\documentclass[12pt,preprint]{aastex}
\usepackage[dvips]{epsfig,color}
\usepackage{amsmath}

\newcommand{\eg}{{\it e.g.,}}
\newcommand{\ie}{{\it i.e.,}}
\newcommand{\etal}{{\it et al.}}

\newcommand{\ignore}[1]{\relax}

\DeclareMathAlphabet{\mathsfsl}{OT1}{cmss}{m}{sl}

\newcommand{\dif}{\mathrm{d}}

\newcommand{\vnab}{\ensuremath{\boldsymbol{\nabla}}}
\newcommand{\vvz}{\ensuremath{\boldsymbol{v_0}}}
\newcommand{\vvp}{\ensuremath{\boldsymbol{v_p}}}

\newcommand{\bcdot}{\ensuremath{\boldsymbol{\cdot}}}
\newcommand{\dfdt}{\ensuremath{\frac{\partial f}{\partial t}}}

\newcommand{\kb}{\ensuremath{k_{\rm B}}}
\newcommand{\vx}{\ensuremath{\boldsymbol{x}}}
\newcommand{\vv}{\ensuremath{\boldsymbol{v}}}
\newcommand{\va}{\ensuremath{\boldsymbol{a}}}

\shorttitle{}
\shortauthors{Barnes \etal}

\begin{document}

\title{Entropy Production in Collisionless Systems. I. Large
Phase-Space Occupation Numbers}
\author{Eric I. Barnes}
\affil{Department of Physics, University of Wisconsin --- La Crosse,
La Crosse, WI 54601}
\email{barnes.eric@uwlax.edu}
\author{Liliya L. R. Williams}
\affil{Department of Astronomy, University of Minnesota,
Minneapolis, MN 55455}
\email{llrw@astro.umn.edu}

\begin{abstract}

Certain thermal non-equilibrium situations, outside of the
astrophysical realm, suggest that entropy production extrema, instead
of entropy extrema, are related to stationary states.  In an effort to
better understand the evolution of collisionless self-gravitating
systems, we investigate the role of entropy production and develop
expressions for the entropy production rate in two particular
statistical families that describe self-gravitating systems. From
these entropy production descriptions, we derive the requirements for
extremizing the entropy production rate in terms of specific forms for
the relaxation function in the Boltzmann equation.  We discuss some
implications of these relaxation functions and point to future work
that will further test this novel thermodynamic viewpoint of
collisionless relaxation.

\end{abstract}

\keywords{galaxies:structure --- galaxies:kinematics and dynamics}

\section{Introduction}\label{intro}

Galaxy-hosting dark matter systems and the galactic stellar systems
themselves act collisionlessly over Hubble timescales.  Understanding
the evolution of self-gravitating, collisionless systems is a
foundational element to the larger picture of galaxy evolution.  Much
of the advancement in understanding collisionless evolution has come
from $N$-body simulations that focus on the motions of large numbers
of individual mass elements.  Such simulations allow astrophysicists
to determine the density and velocity distributions that result from a
variety of initial conditions \citep[\eg][]{va82,ma85,nfw96,m98}.

While $N$-body simulations have given the astrophysics community a
powerful method for making predictions that can be tested against
observations, the modeled evolutions are sufficiently complex that a
full physical picture of the evolution remains lacking.  Effects such
as dynamical friction, collisionless relaxation processes, and even
numerical artifacts can all play roles of varying importance in these
simulations.  Astrophysicists now have substantial empirical evidence
that collisionless systems in cosmological contexts have ``universal''
equilibrium distributions of mass and velocities, \eg\ the radial
power-law behavior of density divided by velocity dispersion cubed
\citep{tn01}.  We are interested in gaining a better understanding of
the physical origin of these seemingly special distributions.

Before beginning to build a physical picture of collisionless
evolution, let us quickly review some of the ideas that are involved.
Thermal equilibrium is the state that systems relax towards when
energy can be exchanged between the components of the system.  A
system in thermal equilibrium is characterized by a minimum internal
energy or, equivalently, a maximum entropy.  In a thermodynamic sense,
entropy is linked to the amount of energy transferred as heat,
bringing with it a connotation of randomness.  In a statistical sense,
entropy is simply a redefinition of the accounting of energy states in
a system.  Relaxation is a generic term for any process that erases
the memory of initial conditions in a system.  For example, gases
reach the completely relaxed state of thermal equilibrium through
collisions.

\citet{lb67} demonstrated that non-degenerate collisionless
self-gravitating systems with finite mass and energy do not have a
state of maximum entropy when one works with the standard distribution
function and assumes that the occupation numbers of phase-space
volumes is large enough that Stirling's approximation applies
\citep[For an alternative view see][]{hw10}.  This lack of entropy
extremum implies that there is no thermal equilibrium state for such
systems.  And yet, mechanical equilibrium, a state with no net force
at any location, is possible.  In stellar interiors, for example, the
lack of thermal equilibrium---evidenced by gradients in temperature
and pressure---is required for mechanical equilibrium.  Note that the
thermal non-equilibrium description refers to the entire star.  It is
common to treat the gas in a star as being in local thermodynamic
equilibrium, where small regions of the star are modeled as having
uniform temperatures.  In this work, we consciously make a distinction
between full thermal equilibrium and stationary (time-independent)
situations.  An example that clarifies this difference is familiar to
those of us nearer the poles.  On a cold winter day, a house is heated
to a temperature that is well above the outside air temperature.  Heat
flows out of the house, and entropy is generated in a wall during the
energy transfer.  As long as the home's heating system runs and the
outdoor temperature remains constant, this process will be
steady-state, yet it does not represent a thermal equilibrium as the
entropy of the home-outside air system continually increases.

Dark matter halos and galactic stellar systems are similar to the above,
in the sense that they have temperature gradients and are thus thermally 
non-equilibrium systems, even though they are steady-state and are in 
mechanical equilibrium. 

Thermal non-equilibrium systems are common in nature and have been the
subject of numerous studies. For certain non-astrophysical,
non-gravitating systems it was found that stationary states, like
mechanical equilibrium, coincide with states of extreme entropy
production \citep[][and references therein]{dgm84,g08}.  In some
instances, it appears that steady-states are reached by minimizing
entropy production, while in other cases maximizing entropy production
is the requirement. The most common modern formulation of the
principle of maximum entropy production is due to \citet{z61}. The
principle of miminum entropy production \citep{p78} appears to
have a smaller range of applicability, but does find uses in physics,
chemistry and biology. The theoretical interpretation, and the
relation between the two seemingly contradictory principles is still
being worked out \citep[\eg,][]{ms06,b10}.

A further, and more general reason to explore entropy production in
self-gravitating systems is because irreversability and entropy
production are closely related concepts. The ``microscopic'' equation
of motion of gravity (and of other microscopic physics) is
time-reversible, while the global evolution of the system is not, in
the sense that it has a definite arrow of time. A natural agent that
can make the transition from reversability to irreversability in
macroscopic systems is the thermodynamic entropy production. 

This ability of thermodynamics to pick the arrow of time through
non-zero (positive) entropy production has been utilized in a
cosmological context by \citet{pg86}.  These authors propose a
scenario where in the early Universe some of the energy of the
space-time (gravity) was transferred to matter. Because matter
production is an irreversible, entropy creating process, the
evolution of the Universe proceeds with an arrow of time.

In this paper we investigate the consequences of applying the entropy
production extremization principle to self-gravitating systems.  We do
not prove that such a principle should be applicable here; we merely
explore what this principle implies. In particular, we want to find
what conditions are necessary to extremize entropy production and how
those conditions differ from maximizing the total entropy. 

We begin with a review of relevant ideas regarding entropy from both a
thermodynamic and statistical point of view in Section~\ref{back}.  In
\S~\ref{kinetic}, entropy production equations for the two statistical
families we are interested in are then derived.  For these cases, we
also develop the specific forms for collisionless relaxation processes
required to extremize the entropy production.  We present a discussion
of our results and conclude in Section~\ref{discuss}.

\section{Background}\label{back}

\subsection{Brief Outline}

Much of what follows is a distilled derivation of the outcome of
extemizing entropy production in collisionless systems.  As the
mathematical details might muddy the overall flow of the paper, we
present an executive summary to draw the reader's attention to the
main equations and the connections between them.

\begin{itemize}

\item We start by developing a continuity equation for entropy density
($\rho s$), which has an entropy production source term $\sigma$ (see
Equation~\ref{s6}).

\item Statistical descriptions of entropy and entropy density are
derived from enumerating states accessible to the system.  Entropy
density expressions for both Maxwell-Boltzmann and Lynden-Bell
statistical families may be found in Equations~\ref{mbsdens} and
\ref{lbsdens}, respectively. These expressions depend on the
coarse-grained distribution function, $f$.

\item Taking the partial time derivative of $\rho s$ in
Equations~\ref{mbsdens} and \ref{lbsdens} produces
Equations~\ref{mbk1} and \ref{lbk1}, respectively.  These
diffentiations result in $\partial f/\partial t$ terms, which we
replace using the Boltzmann Equation (\ref{boltz}). This
introduces the relaxation function $\Gamma$, as shown in 
Equations~\ref{mbk2} and \ref{lbk2}, respectively. 

\item Expressions of the entropy production $\sigma$ for both
statistical families are identified by comparing terms resulting from
Equations~\ref{mbk2} and \ref{lbk2} to those in Equation~\ref{s6}.

\item Finally, we set the variation of the $\sigma$ expressions equal
to zero.  This procedure then yields the $\Gamma$ functions found in
Equations~\ref{mbexts} and \ref{lbexts}, respectively.

\end{itemize}

\subsection{Entropy Behavior}\label{entropy}

Following \citet[][Ch. 3]{dgm84}, we argue that there are two ways to
change the total entropy of a system; 1) entropy can enter or
leave through a flux, or 2) entropy may be created within the system.
Note that to be in accord with the second law of thermodynamics during
the system's evolution, the system's entropy should not decrease.
A straightforward approach to dealing with entropy changes in time is,
\begin{equation}\label{s1}
\frac{\partial S}{\partial t} = \frac{\partial S^{(f)}}{\partial t}
+ \frac{\partial S^{(c)}}{\partial t},
\end{equation}
where the total entropy is,
\begin{equation}\label{s2}
S=\int \rho s \: \dif\vx,
\end{equation}
$S^{(f)}$ is the flux entropy, and $S^{(c)}$ is the created entropy.
Unless noted otherwise, all integrations are assumed to be over all
configuration and/or velocity space.  The value of the flux entropy
depends on how much entropy is entering or leaving the system through
a bounding surface,
\begin{eqnarray}\label{s3}
\frac{\partial S^{(f)}}{\partial t} & = & -\int_{\rm S}
(\boldsymbol{\Sigma} + \rho s \vvz) \bcdot \dif \boldsymbol{A} \\
\nonumber & = & -\int_{\rm V} \vnab \bcdot (\boldsymbol{\Sigma} +
\rho s \vvz) \: \dif\vx,
\end{eqnarray}
where $\boldsymbol{\Sigma}$ is an entropy flux normal to the surface
S, $\rho s \vvz$ is the entropy convected through the surface by mean
motions, the minus sign is convention, and the last equality results
from the divergence theorem.  The flux $\boldsymbol{\Sigma}$ is the
result of random motions within a system.  In as much as random
motions can be linked to thermal energy, one can think of the random
flux as being associated with heat.  The subscripts on the integrals simply
draw attention to the change in integration range.  The entropy
creation term is likewise related to a quantity associated with the
system,
\begin{equation}\label{s4}
\frac{\partial S^{(c)}}{\partial t} = \int \sigma \: \dif\vx,
\end{equation}
where $\sigma$ is the entropy production per unit volume per unit
time.  Substituting Equations~\ref{s2}, \ref{s3}, and \ref{s4} into
Equation~\ref{s1}, 
\begin{equation}\label{s5}
\int \frac{\partial}{\partial t}(\rho s) \: \dif\vx =
-\int \vnab \bcdot
(\boldsymbol{\Sigma} + \rho s \vvz) \: \dif\vx +
\int \sigma \: \dif\vx.
\end{equation}
As this expression must be true for an arbitrary volume,
\begin{equation}\label{s6}
\frac{\partial}{\partial t}(\rho s) = -\vnab \bcdot
(\boldsymbol{\Sigma} + \rho s \vvz) + \sigma.
\end{equation}
This is a continuity equation for the entropy density. The source
term, $\sigma$, represents the entropy production within any
infinitesimal volume of the system. 

\subsection{Statistical Representations}\label{stat}

From a statistical point of view, entropy is simply a measurement of
the number of states accessible to a particular system.  From the
statistical definition of temperature, entropy can be calculated using,
\begin{equation}\label{s0}
S=\kb \ln{\Omega},
\end{equation}
where $\Omega$ is the number of accessible states and \kb\ is the
Boltzmann constant which serves to give entropy the correct
thermodynamic units.  As a result, counting procedures are key to
determining specific realizations of entropy.  \citet{lb67} discusses
how there are four counting styles that lead to physically relevant
situations.  Bose-Einstein statistics follow from counting states for
indistinguishable particles that can co-habitate.  When
indistinguishable particles are not allowed to share states,
Fermi-Dirac statistics emerge.  Classical Maxwell-Boltzmann (MB)
statistics result from counting states available to distinguishable
particles that can share states.  Completing the symmetry, systems
where distinguishable particles cannot co-occupy states obey what has
become known as Lynden-Bell (LB) statistics (statistics `IV' in
Lynden-Bell's original notation).

Each type of statistics will produce different representations of
entropy, but we will focus on MB and LB here, both of which deal with
classical distinguishable particles.  We briefly recap the notation
used in \citet{lb67} before proceeding with our discussion.  The usual
six-dimensional position-velocity phase-space (\vx,\vv) is the setting
for determining the statistics.  Imagine phase-space to be divided
into a very large number of nearly infinitesimal volumes $\varpi$
called micro-cells.  Each micro-cell can either be occupied or
unoccupied by one of the $N$ phase elements of the system.  These
phase elements can be thought of as being represented by the
fine-grained distribution function.  This fine-grained distribution
function is assumed to have a constant value $\eta=1/\varpi$.  At this
level, the phase elements cannot co-habitate.  If phase-space is also
partitioned on a coarser level so that some number of micro-cells
$\nu$ occupy a macro-cell, then we can discuss a coarse-grained
distribution function.  The volume of a macro-cell is then $\nu\varpi$
and the $i$th macro-cell contains $n_i$ phase elements.  We assume
that while the volume of a macro-cell is much larger than that of a
micro-cell, it is still very small compared to the full extent of
phase-space occupied by the system.  The number of ways of organizing
the $n_i$ elements into the $\nu$ micro-cells without co-habitation
is,
\begin{equation}\label{lbw0}
\frac{\nu!}{(\nu-n_i)!}.
\end{equation}
If the elements were allowed to multiply occupy micro-cells (as in MB
statistics) this number would be $\nu^{n_i}$.

To get the total number of accessible states, the possible ways to
distribute the $N$ phase elements into $n_i$ chunks must also be
included.  \citet{lb67} derives,
\begin{equation}\label{lbw}
\Omega_{\rm LB} = \frac{N!}{\prod_i n_i!} \times \prod_i 
\frac{\nu!}{(\nu-n_i)!}.
\end{equation}
In the MB case, the only change is that the factorial ratio in the final
product term is replaced by $\nu^{n_i}$.  With Equation~\ref{lbw}, the
entropy can now be expressed as,
\begin{eqnarray}\label{lbs0}
S_{\rm LB} & = & \kb \left\{ \ln{N!} - \sum_i \ln{n_i!} + \sum_i
\ln{\nu!} - \sum_i \ln{(\nu-n_i)!} \right\} \nonumber \\
 & = & \kb \left\{ N(\ln{N}-1) + \sum_i \nu\ln{\nu} -
\sum_i [ n_i(\ln{n_i}-1) + (\nu-n_i)\ln{(\nu-n_i)} ] \right\},
\end{eqnarray}
where we have assumed $N \gg 1$, $n_i \gg 1$, $\nu \gg 1$, and used
Stirling's approximation.  In a Maxwell-Boltzmann situation, the
familiar entropy expression is,
\begin{equation}\label{mbs0}
S_{\rm MB}= \mathcal{S}_0 - \kb \sum_i n_i\ln{n_i},
\end{equation}
where $\mathcal{S}_0 = N\kb \ln{N\nu}$.  Note that $S_{\rm LB}$
reduces to an expression similar to $S_{\rm MB}$ (except for the
constant) when $\nu \gg n_i$.  In this dilute, non-degenerate limit,
the difference between allowing and denying co-habitation of states
becomes irrelevant.  The number of elements occupying a macro-cell
$n_i$ is now replaced by a coarse-grained distribution function of
position and velocity.  The coarse-grained function $f$ is a
continuous function of \vx\ and \vv\ and represents the number of
occupied micro-cells within macro-cells,
\begin{equation}
N=\sum_i n_i=\iint f \: \dif\vx \dif\vv,
\end{equation}
where the integration is taken over all of phase-space.

\citet{lb67} goes on to investigate the results of minimizing
Equation~\ref{lbs0} with the constraints of constant system mass and
energy.  The well-know result of this work is that non-degenerate
self-gravitating systems do not have a finite mass and energy
thermodynamic equilibrium.  

We will now describe our investigation into the entropy production
that occurs in these systems, keeping the assumption that Stirling's
approximation is accurate.  Note that entropy production takes
place because we are considering coarse-grained distributions; in
other words, particles are now allowed to occupy states not accessible
within the fine-grained world view.  At the fine-grained level the
distribution function does not change and there is no change in
entropy.   

\section{Kinetic Theory Picture}\label{kinetic}

We now proceed to combine the fluid picture of entropy presented in
Section~\ref{entropy} with the statistical picture from
Section~\ref{stat}.  The general procedure will be to find the time
rate of change of the statistical versions of entropy and equate terms
to those in Equation~\ref{s6}.  To do this, we will rely on the
Boltzmann equation.  Note that the coarse-grained distribution
function $f$ does not obey the collisionless Boltzmann equation,
rather its evolution is defined by,
\begin{equation}\label{boltz}
\dfdt = -\boldsymbol{v} \bcdot \vnab f -
\va \bcdot \vnab_v f + \Gamma(f),
\end{equation}
where \va\ is the acceleration and $\vnab_v$ is the velocity space
gradient operator ($\vnab_v=\partial/\partial v_x \hat{i} +
\partial/\partial v_y \hat{j} + \partial/\partial v_z \hat{k}$).  In
standard, ideal gas-like situations $\Gamma$ describes how
inter-particle, two-body collisions can affect the number of particles
in a given phase-space volume.  In the systems that we are dealing
with, $\Gamma$ represents the effect of collisionless relaxation
processes, like violent relaxation and phase mixing, which may be 
aspects of Landau damping \citep{k98}.  We will refer to $\Gamma$
as the relaxation function in subsequent sections.

\subsection{Maxwell-Boltzmann Statistics}\label{mb}

As a simple introduction to the procedure, we begin by discussing a
system in the Maxwell-Boltzmann regime.  Discussion of Lynden-Bell-type 
systems will follow in Section~\ref{lb}. From the entropy form given in
Equation~\ref{mbs0}, our goal is to transform the summation over particles
into an integration over phase space.  First, we re-write
Equation~\ref{mbs0} as
\begin{equation}
S_{\rm MB} = -\kb \sum_i n_i \left(\ln{n_i} -
\frac{\mathcal{S}_0}{N\kb} \right),
\end{equation}
by taking the constant term into the summation.  Next, we use the fact
that $n_i=(\nu/\eta) f$, where $f=f(\vx,\vv)$, to produce,
\begin{equation}\label{mbs1}
S_{\rm MB}=-\frac{\kb}{\varpi} \iint \frac{f}{\eta} \left[ 
\ln{\left( \frac{f}{\eta}\right)} - \ln{N} \right] \: 
\dif\vx \dif\vv.
\end{equation}
The entropy density can now be written as,
\begin{equation}\label{mbsdens}
\rho s_{\rm MB} = -\frac{\kb}{\varpi} \int \frac{f}{\eta} \left[ 
\ln{\left( \frac{f}{\eta}\right)} - \ln{N} \right] \: \dif\vv.
\end{equation}
Note that this expression for entropy density differs slightly from
other kinetic theory derivations in the constant term ($\ln{N}$ versus
1).  This is due to a different choice for the additive constant in
the definition of entropy.  As such, it does not affect the behavior
we are going to investigate.

Taking a partial time derivative of Equation~\ref{mbsdens} results in,
\begin{eqnarray}\label{mbk1}
\frac{\partial}{\partial t}(\rho s_{\rm MB}) & = & -\kb
\int \dfdt \left[ \ln{\left(\frac{f}{\eta}\right)} + 1 - \ln{N}
\right] \: \dif\vv.
\end{eqnarray}

Substituting $\partial f/\partial t$ from the Boltzmann equation into
Equation~\ref{mbk1} results in a lengthy expression that we will deal
with term by term.  For reference, the expression after substitution
is,
\begin{eqnarray}\label{mbk2}
\frac{\partial}{\partial t}(\rho s_{\rm MB}) & = & 
-\kb \int (-\vv \bcdot \vnab f) \left[
\ln{\left(\frac{f}{\eta}\right)} +1 - \ln{N} \right] \: \dif\vv -
\nonumber \\
 & & \kb \int (-\va \bcdot \vnab_v f) \left[
\ln{\left(\frac{f}{\eta}\right)} +1 - \ln{N} \right] \: \dif\vv - 
\nonumber \\ 
& & \kb \int \Gamma \left[ 
\ln{\left(\frac{f}{\eta}\right)} +1 - \ln{N} \right] \: \dif\vv.
\end{eqnarray}

The first integral on the right-hand side of Equation~\ref{mbk2} can be
recast using integration by parts and the fact that $\vv \bcdot \vnab
f = \vnab \bcdot (f\vv)$ in phase-space,
\begin{equation}\label{mbk3}
\int (\vv \bcdot \vnab f) \left[ \ln{\left(\frac{f}{\eta}\right)} +1 -
\ln{N} \right] \: \dif\vv
= \vnab \bcdot \int f \left( \ln{\left(\frac{f}{\eta}\right)}
- \ln{N} \right) \vv \: \dif\vv.
\end{equation}
Since \va\ is velocity-independent, $\va \bcdot \vnab_v f = \vnab_v
\bcdot \va f$, and we can again use integration by parts to transform
the second term on the right-hand side of Equation~\ref{mbk2} to,
\begin{equation}
\int \vnab_v \bcdot \left[ \va f \ln{\left(\frac{f}{\eta}\right)}
\right] \: \dif\vv - \ln{N}\int \va \bcdot \vnab_v f \: \dif\vv.
\end{equation}
Using the divergence theorem and the fact that any physical
distribution function must vanish for large velocities, these
integrals vanish.  Equation~\ref{mbk2} now has the form,
\begin{equation}\label{mbk4}
\frac{\partial}{\partial t}(\rho s_{\rm MB}) = 
-\kb \left\{ -\vnab \bcdot \int f \left(
\ln{\left(\frac{f}{\eta}\right)} - \ln{N} \right) \vv \: \dif\vv +
\int \Gamma \left[ \ln{\left(\frac{f}{\eta}\right)} +1 - \ln{N}
\right] \right\} \: \dif\vv.
\end{equation}
 
Assuming the velocity field to be composed of mean and
peculiar components $\vv = \vvz + \vvp$, we can re-cast
Equation~\ref{mbk4} as,
\begin{eqnarray}\label{mbk5}
\frac{\partial}{\partial t}(\rho s_{\rm MB}) & =  &
-\vnab \bcdot \rho s_{\rm MB} \vvz -\kb \Bigg\{ 
-\vnab \bcdot \int f \left(
\ln{\left(\frac{f}{\eta}\right)} - \ln{N} \right) \vvp \: \dif\vv +
\nonumber \\
 & & \int \Gamma \left[ \ln{\left(\frac{f}{\eta}\right)} +1 - \ln{N}
\right] \: \dif\vv \Bigg\}.
\end{eqnarray}
We now equate the terms in Equation~\ref{mbk5} to those in
Equation~\ref{s6},
\begin{equation}
\frac{\partial}{\partial t}(\rho s) = -\vnab \bcdot
(\boldsymbol{\Sigma} + \rho s \vvz) + \sigma.
\end{equation}
The entropy flux $\boldsymbol{\Sigma}$ is given by the integral in the
second term on the right-hand side of Equation~\ref{mbk5} and
represents randomly fluxed entropy.
The remaining term is the entropy production for the system,
\begin{equation}\label{mbsprod}
\sigma_{\rm MB}=
-\kb \int \Gamma \left[ 
\ln{\left(\frac{f}{\eta}\right)} +1 -\ln{N} \right] \: \dif\vv.
\end{equation}
This equation explicitly demonstrates how the non-collisionless nature
of the coarse-grained distribution function leads to changes in
entropy.

\begin{figure}
\plotone{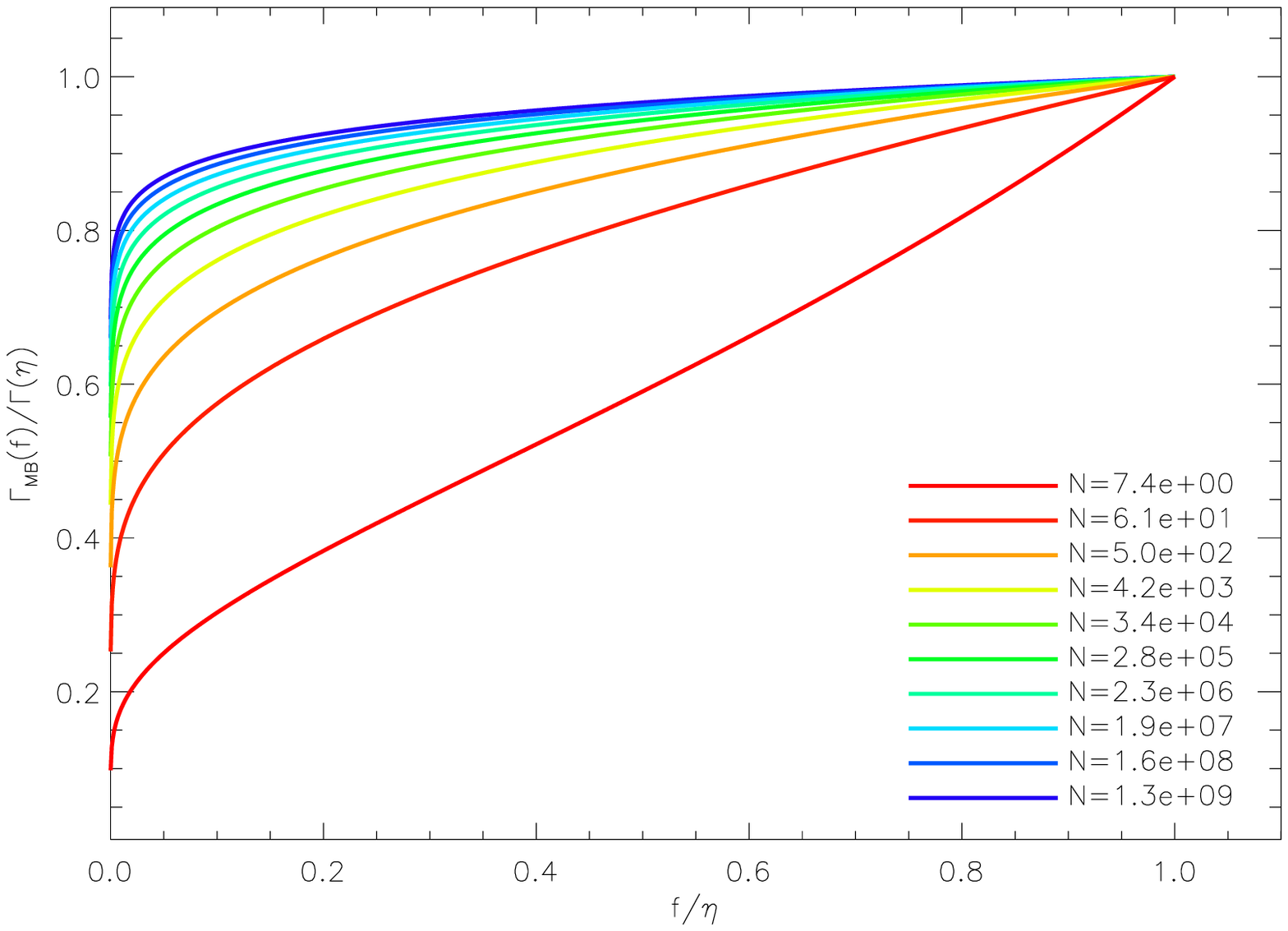}
\figcaption{The behavior of the relaxation function $\Gamma_{\rm MB}$
as a function of the coarse-grained distribution function $f$ in a
system that obeys Maxwell-Boltzmann statistics.  The various lines
represent functions defined with different $N$ values, where $N$ is
basically the number of particles in a system.  As $N$ increases, the
relaxation function becomes more constant.
\label{mbgplot}}
\end{figure}

As mentioned in \S~\ref{intro}, thermodynamic non-equilibrium systems
can have steady states described by extrema of entropy production.
Note that we do not extremize the entropy itself; that procedure
implies a system in thermodynamic equilibrium \citep[but not
necessarily with constant kinetic temperature, see][]{hw10}.  What we
wish to do is to set, $\delta \sigma_{\rm MB} = 0$.  Taking the
variation of Equation~\ref{mbsprod}, we find
\begin{equation}
\delta \sigma_{\rm MB} = -\kb \int \delta f \left\{
\frac{\dif \Gamma}{\dif f} \left[ \ln{\left(\frac{f}{\eta}\right)} +1
- \ln{N} \right] + \frac{\Gamma}{f} \right\} \: \dif\vv.
\end{equation}
Since $\delta f$ is arbitrary, the variation disappears only when the
term in curly braces is zero. Note that we did not impose any additional
constraints while carrying out the extremization. In the standard entropy
extremization procedure Langrange multiplier terms are introduced to assure
constant mass and energy. The condition for an extremum in
entropy production is,
\begin{equation}
\frac{\dif \Gamma}{\dif f} \left( \ln{\left(\frac{f}{\eta}\right)} +1
- \ln{N} \right) + \frac{\Gamma}{f} =
\frac{\dif \Gamma}{\dif (f/\eta)} \left( \ln{\left(\frac{f}{\eta}\right)} +1
- \ln{N} \right) + \frac{\Gamma}{(f/\eta)} = 0.
\end{equation}
This can be rewritten as,
\begin{equation}
\frac{\dif \ln{\Gamma}}{\dif \ln{(f/\eta)}} = -\frac{1}{\ln{(f/\eta)}
 + 1 - \ln{N}},
\end{equation}
with the following solution for the relaxation function,
\begin{equation}\label{mbexts}
\Gamma_{\rm MB}(f)=\frac{(1-\ln{N})\Gamma_{\rm MB}(f=\eta)}
{\ln{(f/\eta)} +1 -\ln{N}},
\end{equation}
where the value of the fine-grained distribution function $\eta$ is
the maximum value allowed for $f$.  In a collisional system that
follows the Maxwell-Boltzmann distribution, like an ordinary gas, one
could imagine $f$ reaching any value due to collisions.  However, for
the collisionless case, MB statistics are the non-degenerate limit,
implying that $f \ll \eta$.  Figure~\ref{mbgplot} contains plots of
$\Gamma_{\rm MB}(f)$ for a variety of values of $N$.  As $N$
increases, the relaxation function becomes more and more uniform.

\subsection{Lynden-Bell Statistics}\label{lb}

We now investigate a system that obeys Lynden-Bell statistics.  The
entropy of the system can be transformed from the summation in
Equation~\ref{lbs0} to,
\begin{equation}\label{lbs1}
S_{\rm LB}=-\kb \iint f \left[ 
\ln{\left( \frac{\nu}{\eta}f\right)} - 
\frac{S_0}{N\kb} \right] + (\eta-f)\ln{\left( \frac{\nu}{\eta}(\eta-f)
\right)} \: \dif\vx \dif\vv,
\end{equation}
where $S_0=N\kb(\ln{N}-1)+\kb\sum_i \nu\ln{\nu}$.
The entropy density can now be written as,
\begin{equation}\label{lbsdens}
\rho s_{\rm LB} = -\kb \int f \left[ 
\ln{\left( \frac{\nu}{\eta}f\right)} - 
\frac{S_0}{N\kb} \right] + (\eta-f)\ln{\left( \frac{\nu}{\eta}(\eta-f)
\right)} \: \dif\vv.
\end{equation}

Taking a partial time derivative of Equation~\ref{lbsdens} results in
\begin{eqnarray}\label{lbk1}
\frac{\partial}{\partial t}(\rho s_{\rm LB}) & = & -\kb
\int \dfdt
\left[ \ln{\left(\frac{f}{\eta-f}\right)} -C \right] \: \dif\vv
\nonumber \\ & = & -\kb \int \dfdt
\ln{\left(\frac{f}{\eta-f}\right)} \: \dif\vv + \kb C \int \dfdt \:
\dif\vv,
\end{eqnarray}
where $C=S_0/N\kb$ is a constant.

Substituting $\partial f/\partial t$ from the Boltzmann equation into
Equation~\ref{lbk1} again results in a lengthy expression that we will
deal with term by term.  For reference, the expression after
substitution is,
\begin{eqnarray}\label{lbk2}
\lefteqn{
\frac{\partial}{\partial t}(\rho s_{\rm LB}) = } \nonumber \\
& & -\kb \int (-\vv \bcdot \vnab f) 
\ln{\left(\frac{f}{\eta-f}\right)}
\: \dif\vv -\kb \int (-\va \bcdot \vnab_v f) 
\ln{\left(\frac{f}{\eta-f}\right)} \: \dif\vv - \nonumber \\ 
& & \kb \int \Gamma 
\ln{\left(\frac{f}{\eta-f}\right)} \: \dif\vv + 
\kb \int (-\vv \bcdot \vnab f)C \: \dif\vv + 
\kb \int (-\va \bcdot \vnab_v f)C \: \dif\vv + 
\nonumber \\
& & \kb \int \Gamma C \: \dif\vv.
\end{eqnarray}

The first integral on the right-hand side of Equation~\ref{lbk2} can
be transformed to,
\begin{equation}\label{lbk3}
\int (\vv \bcdot \vnab f)\ln{\left(\frac{f}{\eta-f}\right)}\: \dif\vv
= \vnab \bcdot \int f\ln{\left(\frac{f}{\eta-f}\right)} \vv \:
\dif\vv - \int \frac{\eta}{\eta-f} \vv \bcdot \vnab f \: \dif\vv.
\end{equation}
We now re-write the first term on the right-hand side of
Equation~\ref{lbk3} as,
\begin{equation}\label{lbk4}
\vnab \bcdot \int f\ln{\left(\frac{f}{\eta-f}\right)} \vv \: \dif\vv
= \vnab \bcdot \int f \left[ \ln{\left(\frac{\nu}{\eta}f\right)} -
\ln{\left(\frac{\nu}{\eta}(\eta-f)\right)} \right] \vv \: \dif\vv.
\end{equation}
Adding and subtracting terms of the form $\vnab \bcdot \int 
\eta \ln{[(\nu/\eta)(\eta-f)]} \vv \: \dif\vv$ to Equation~\ref{lbk4}
gives us a term that is very reminiscent of the entropy density in
Equation~\ref{lbsdens},
\begin{eqnarray}\label{lbk5}
\lefteqn{
\int (\vv \bcdot \vnab f)\ln{\left(\frac{f}{\eta-f}\right)}\: \dif\vv
=} \nonumber \\
& & \vnab \bcdot \int \left[ f \ln{\left(\frac{\nu}{\eta}f\right)} + 
(\eta-f) \ln{\left(\frac{\nu}{\eta}(\eta-f)\right)} \right] \vv \: 
\dif\vv - \nonumber \\
& & \vnab \bcdot \int \eta \ln{\left(\frac{\nu}{\eta}(\eta-f) \right)} 
\vv \: \dif\vv - \int \frac{\eta}{\eta-f} \vv \bcdot \vnab f \: 
\dif\vv.
\end{eqnarray}
Looking at the second term on the right-hand side of
Equation~\ref{lbk5}, we see that
\begin{equation}
\vnab \bcdot \eta\ln{\left( \frac{\nu}{\eta}(\eta-f)\right)}\vv =
-\frac{\eta}{\eta-f} \vv \bcdot \vnab f.
\end{equation}
We can now simplify Equation~\ref{lbk5} to
\begin{equation}\label{lbk6}
\int (\vv \bcdot \vnab f)\ln{\left(\frac{f}{\eta-f}\right)}\: \dif\vv
= \vnab \bcdot \int \left[ f \ln{\left(\frac{\nu}{\eta}f\right)} + 
(\eta-f) \ln{\left(\frac{\nu}{\eta}(\eta-f)\right)} \right] \vv \: 
\dif\vv.
\end{equation}
 
Next, we deal with the second integral on the right-hand side of
Equation~\ref{lbk2}.  Since $\va$ is velocity-independent, we can
write
\begin{equation}\label{lbkk1}
\int (\va \bcdot \vnab_v f) \ln{\left(\frac{f}{\eta-f}\right)} \:
\dif\vv = \va \bcdot \int (\vnab_v f) 
\ln{\left(\frac{f}{\eta-f}\right)} \: \dif\vv.
\end{equation}
Using
\begin{equation}
(\vnab_v f) \ln{\left(\frac{f}{\eta-f}\right)} =
\vnab_v \left[ f \ln{\left(\frac{f}{\eta-f}\right)} \right] -
\frac{\eta}{\eta-f}\vnab_v f,
\end{equation}
and taking advantage of the velocity independence of \va,
we can re-write the right-hand side of Equation~\ref{lbkk1} as,
\begin{equation}\label{lbkk2}
\int \vnab_v \bcdot \left[ \va f \ln{\left(\frac{f}{\eta-f}\right)}
\right]
 \: \dif\vv - \int \frac{\eta}{\eta-f} \va \bcdot \vnab_v f \:
\dif\vv.
\end{equation}
The first term disappears when the divergence theorem is applied, as
the distribution function goes to zero at large velocities.  To deal
with the second term, we substitute $\vnab_v f= -\vnab_v (\eta - f)$
to get,
\begin{equation}
-\int \frac{\eta}{\eta-f} \va \bcdot \vnab_v f \: \dif\vv = \eta \va
\bcdot \int \vnab_v \ln{(\eta-f)} \: \dif\vv.
\end{equation}
Let us investigate one component of this vector integral,
\begin{equation}
\int_{-\infty}^{\infty} \frac{\partial}{\partial v_x} \ln{(\eta-f)} \:
\dif v_x = \ln{(\eta-f)} \big|_{-v_{\rm max}}^{v_{\rm max}} =
\ln{\eta} - \ln{\eta} = 0,
\end{equation}
since a well-behaved distribution function is expected to disappear at
the largest speeds possible.  So, the second integral on the
right-hand side of Equation~\ref{lbk2} is zero.

\begin{figure}
\epsscale{0.85}
\plotone{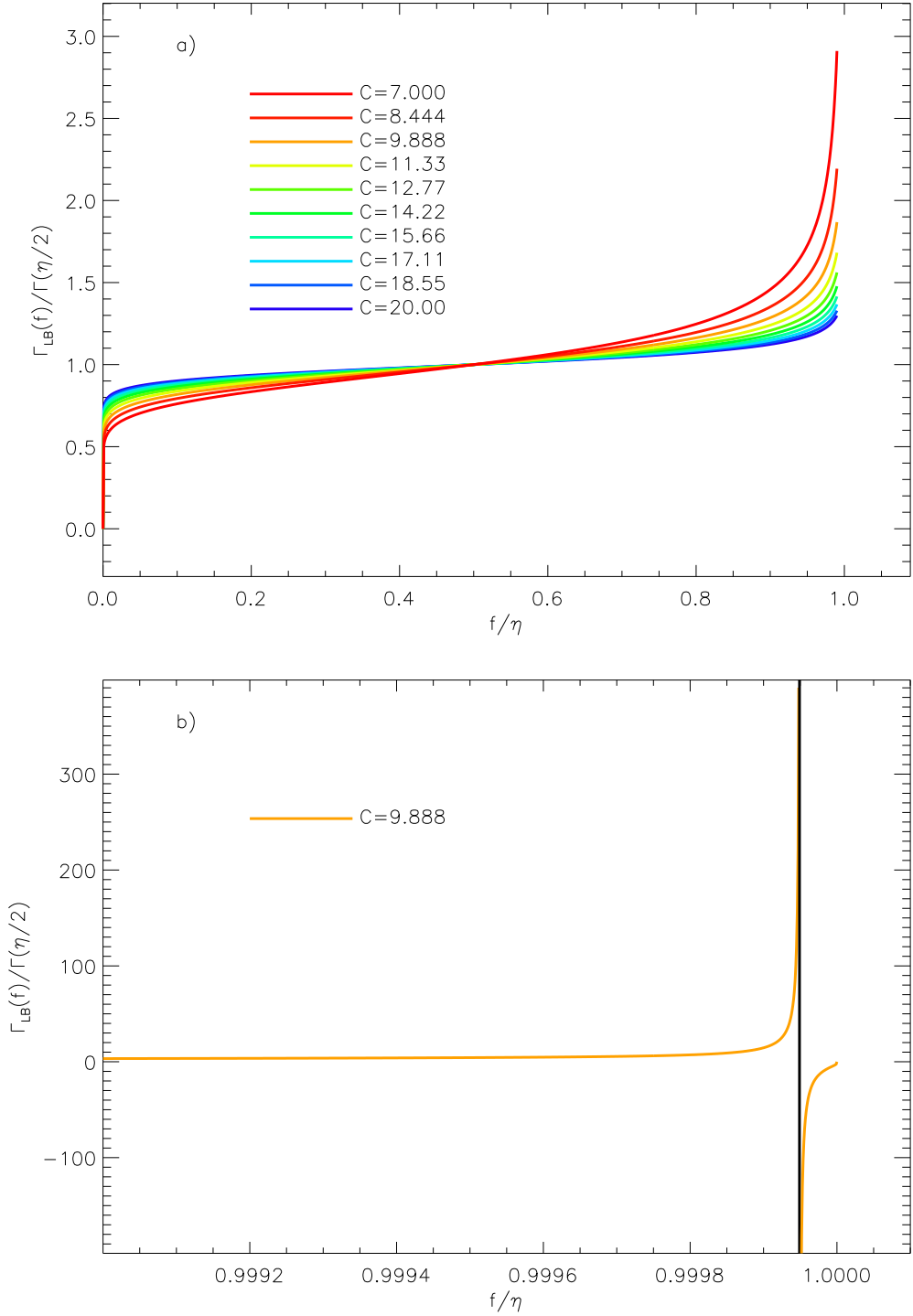}
\figcaption{a) The behavior of the relaxation function $\Gamma_{\rm
LB}$ as a function of the coarse-grained distribution function $f$ in
a system that obeys Lynden-Bell statistics.  The various lines
represent functions defined with different $C$ values, where
$C=\ln{N}-1 + 1/N\sum_i \nu \ln{\nu}$.  As before, $N$ is basically
the number of particles in a system, and increasing $N$ results in
increasing $C$.  Like the Maxwell-Boltzmann case, as $N$ increases,
the relaxation function becomes more constant.  b) A magnified view of
the behavior of $\Gamma_{\rm LB}$ (for a fixed $C$) as $f \rightarrow
\eta$.  Note that $\Gamma_{\rm LB}$ becomes unbound as
$\ln{[f/(\eta-f)]} \rightarrow C$, but as $f \rightarrow \eta$,
$\Gamma_{\rm LB} \rightarrow 0$.
\label{lbgplot}}
\end{figure}

We now re-combine the terms on the right-hand side of
Equation~\ref{lbk2} using the results of Equations~\ref{lbk6} and
\ref{lbkk1} to produce our final representation of
the time rate of change of the entropy density,
\begin{eqnarray}\label{lbkk3}
\lefteqn{
\frac{\partial}{\partial t}(\rho s_{\rm LB}) = } \nonumber \\
& & -\kb \vnab \bcdot \int \left\{ f \left[ 
\ln{\left(\frac{f}{\eta-f}\right)} -C \right] + (\eta-f)\ln{\left(
\frac{\nu}{\eta}(\eta-f)\right)} \right\} \vv \: \dif\vv - \nonumber
\\
& & \kb \int \Gamma 
\left[\ln{\left(\frac{f}{\eta-f}\right)}-C \right] \: \dif\vv.
\end{eqnarray}
Again assuming the velocity field to be composed of mean and
peculiar components $\vv = \vvz + \vvp$, we draw a correspondence
between,
\begin{eqnarray}\label{lbkk4}
\lefteqn{
\frac{\partial}{\partial t}(\rho s_{\rm LB}) = 
-\vnab \bcdot \rho s_{\rm LB} \vvz - } \nonumber \\
& & \kb \vnab \bcdot \int \left\{ f \left[ 
\ln{\left(\frac{f}{\eta-f}\right)} -C \right] + (\eta-f)\ln{\left(
\frac{\nu}{\eta}(\eta-f)\right)} \right\} \vvp \: \dif\vv - \nonumber
\\
& & \kb \int \Gamma 
\left[\ln{\left(\frac{f}{\eta-f}\right)}-C \right] \: \dif\vv,
\end{eqnarray}
and Equation~\ref{s6}
\begin{equation}
\frac{\partial}{\partial t}(\rho s) = -\vnab \bcdot
(\boldsymbol{\Sigma} + \rho s \vvz) + \sigma.
\end{equation}
The entropy flux due to random motions $\boldsymbol{\Sigma}$ is given
by the integral in the second term on the right-hand side of
Equation~\ref{lbkk4}.  The remaining term then makes up the entropy
production for the system,
\begin{equation}\label{lbsprod}
\sigma_{\rm LB}=
-\kb \int \Gamma
\left[\ln{\left(\frac{f}{\eta-f}\right)}-C \right] \: \dif\vv.
\end{equation}
Again, there is a relaxation function-dependent term in the entropy
production.  Following the same steps as in the Maxwell-Boltzmann
case, we find that the condition for the entropy production term in
Equation~\ref{lbsprod} to be extremized is,
\begin{equation}
\Gamma_{\rm LB}(f)=\frac{Q}{\ln{[f/(\eta-f)]}-C},
\end{equation}
where $Q$ is an integration constant and $C=\ln{N}-1 + 1/N \sum_i \nu
\ln{\nu}$.  If we set $f=\eta/2$, $Q=-C \Gamma_{\rm LB}(f=\eta/2)$ and
\begin{equation}\label{lbexts}
\Gamma_{\rm LB}(f)=\frac{-C\Gamma_{\rm LB}(f=\eta/2)}
{\ln{[f/(\eta-f)]}-C}.
\end{equation}
Like $\Gamma_{\rm MB}$, this function behaves more like a constant as
the $N$ value increases ($C \rightarrow \infty$).  $\Gamma_{\rm LB}$
curves for a variety of $C$ values are presented in
Figure~\ref{lbgplot}a.  The increase in $\Gamma_{\rm LB}$ seen as
$f/\eta \rightarrow 1$ is due to $\ln{[f/(\eta-f)]}$ approaching $C$.
Once $f$ is sufficiently close to $\eta$ that subtracting $C$ produces
a negligible change, $\Gamma_{\rm LB} \rightarrow 0$.  The complex
behavior of $\Gamma_{\rm LB}$ for $f \approx \eta$ is shown in
Figure~\ref{lbgplot}b and is discussed further in the next section.

\section{Discussion \& Conclusions}\label{discuss}

The seminal paper by \cite{lb67} has shown that, under certain
assumptions, there is no maximum entropy state for self-gravitating
systems.  Alternative descriptions of the same systems which yield
self-consistent results under entropy extremization may be found in
\citet{m96} and \citet{hw10}.

Lynden-Bell's conclusion is exemplified by the order-of-magnitude
calculation in \citet[][S.\ 4.7]{t86,bt87}, using Maxwell-Boltzmann
statistics.  The result of this calculation is that when a
collisionless system rearranges its mass to become more centrally
concentrated in the core, the entropy of the contracting core mass
decreases. The outer envelope, expanding in response to the
contracting core to satisfy the virial theorem, should have an increase in
entropy.  As a result, the entropy for the entire system increases.
Since a system can always increase its entropy by contracting the core
and expanding the envelope, there is no maximum entropy state, and
hence maximizing entropy will not lead to a satisfactory description
of a steady state. 

Yet, we know from numerous high-resolution $N$-body simulations that
long-lived steady states do exist.  How does one find these
theoretically?  Apparently, one has to resort to means other other
than entropy extremization.  In this paper we try one alternative
approach. We apply a principle of extremizing entropy production rate
to self-gravitating systems.  This principle has been used widely to
describe thermal non-equilibrium, but not in systems that are
self-gravitating.

Our basic hypothesis is that a steady state is obtained by extremizing
the entropy production.  We present expressions for the entropy
production rates for two types of statistics, Maxwell-Boltzmann (MB)
and Lynden-Bell (LB), as Equations~\ref{mbsprod} and \ref{lbsprod},
respectively.  We then find expressions (Equations~\ref{mbexts} and
\ref{lbexts}) for the relaxation term that forms the right hand side
of the coarse-grained Boltzmann equation.  The meaning of these
expressions and the interpretation of our results, under the
assumption that this idea is applicable to self-gravitating systems,
are discussed below. 

\subsection{Entropy production}

The development of the expressions for entropy production rates
$\sigma$ is a central result of this work.  The descriptions of
$\sigma$ for the MB and LB statistical families are given in
Equations~\ref{mbsprod} and \ref{lbsprod}, respectively.  Since we are
dealing with collisionless systems exclusively, one might expect these
to be zero.  In fact, if we were considering the fine-grained
distribution function, there would be no entropy change, no
thermodynamic evolution, as the collisionless $\dif f/\dif t=0$
Boltzmann equation would apply. 

However, we are considering the coarse-grained function. As a system
evolves, the fine-grained distribution function becomes stretched and
twisted in phase-space.  Because the coarse-grained function averages
the fine-grained function with nearby empty regions of phase-space,
the coarse-grained function changes as the system evolves. Now, recall
that entropy represents the number of accessible states.  On the level
of the fine-grained function, the number of accessible states stays
the same.  However, going from a fine-grained to coarse-grained
description implies that there are now regions of phase-space not
occupied by the fine-grained function that are accessible to the
coarse-grained function. This implies that there are more microscopic
ways of realizing a given macroscopic state, leading to more possible
states and larger entropy.  Thus, coarse-graining an evolving
system results in entropy production even in a collisionless system.
In terms of physical processes, the evolution is due to the
larger-scale phase-space evolution of the system driven by
collisionless relaxation processes, like violent relaxation and phase
mixing.  

The above argues that entropy production takes places during evolution
of collisionless systems. But our analysis shows that entropy
production takes place even during the steady state.  Let us start by
writing down the expression for the entropy production during the
steady-state by combining Equations~\ref{mbsprod} and \ref{mbexts} for
the MB case,
\begin{eqnarray}\label{siglast}
\sigma_{\rm MB} & = & -\kb \int Q \Gamma_{\rm
MB}(\eta) \: \dif\vv \nonumber \\
 & = & -\kb Q\Gamma_{\rm MB}(f=\eta)V_{\rm velocity},
\end{eqnarray}
where $V_{\rm velocity}$ is the volume of occupied velocity space.
This is consistent with the second law of thermodynamics since $Q<0$
and all other terms are positive.  A similar expression can be found
for the LB case, using Equations~\ref{lbsprod} and \ref{lbexts}.
Since a steady state is described by an unchanging value of $\sigma$,
any non-zero value of $\sigma$ persists even when a system has reached
mechanical equilibrium.

It is interesting to think of the source of this continued entropy
production.  After a system stops evolving on the macroscopic scale,
it still continues to evolve on ever decreasing microscopic scales as
the fine-grained function continues to stretch and twist almost
indefinitely. The corresponding continual coarse-graining of the ever
evolving fine-grained function on smaller and smaller scales, results
in constant, non-zero entropy production.

\subsection{Interpreting the Boltzmann Equation}

Our expression for the Boltzmann equation states that the relaxation
function, $\Gamma$, determines the rate of change of the
coarse-grained distribution function,
\begin{equation}
\frac{\dif f}{\dif t} = \Gamma(f).
\label{boltzfull}
\end{equation}
Despite the right hand side being non-zero (given by
Equations~\ref{mbexts} and \ref{lbexts} for the MB and LB statistics,
respectively), the above equation does not contradict the assumption of
a stationary state.  A stationary, or steady, state is the Eulerian
viewpoint, \ie\ $\partial f/\partial t = 0$,  while the Boltzmann
equation above is a Lagrangian viewpoint. $\partial f/\partial t = 0$
does not imply $\dif f/\dif t = 0$.  In other words, the relaxation
function $\Gamma$ does not determine the explicit time-dependence of
$f$, which must be zero for stationary states, but rather describes a
flux of occupied cells through phase-space, as do the velocity-driven
($\vv \bcdot \vnab f$) and acceleration-driven ($\va \bcdot \vnab_v
f$) flux terms (c.f. Equation~\ref{boltz}). In the context of the
Lagrangian derivative, we can think of $t$ simply as a parameter that
indicates the location along a particle's or cell's path through
phase-space. 

In collisional systems, the relaxation function is called the
collision term and is usually dealt with in a Fokker-Planck
approximation scheme. In these systems the particles, through two-body
encounters, gradually disperse over the whole available phase-space,
and so $\dif f/\dif t$ following any given particle in an evolving
system does not stay constant, but generally decreases with time.  In
a general collisionless system, this term represents the processes
like violent relaxation and phase mixing, on the coarse-grained scale.
In a steady-state collisionless system, the large scale processes like
violent relaxation no longer operate and the only changes happen on
microscopic scales.  In this context, the left hand side of
Equation~\ref{boltzfull} describes how a particle, or a cell moves
through the system (and $t$ is the parameter). Therefore it is not
unexpected that over some portions of its motion the coarse-grained
density around it will be increasing and over others, it will be
decreasing.  For the MB case, $\Gamma_{\rm MB} > 0$ for $f>0$,
implying that $f$ should continually grow.  However, this is
impossible as the coarse-grained distribution function is limited to a
maximum value $\eta$.  This contradiction arises as MB statistics are
valid only when $f \ll \eta$ so that macro-cells are not multiply
occupied.  On the other hand, the LB case does not present any
contradictions with the Boltzmann equation.  $\Gamma_{\rm LB}$ is zero
when $f=0$ and $f=\eta$, and is positive over the vast majority of the
intervening range.  This behavior guarantees that when the
coarse-grained density reaches its maximum value, the relaxation term
disappears and the system behaves collisionlessly even at the
coarse-grained level.


To sum up, if extremizing entropy production in self-gravitating
systems does lead to steady-state configurations, then
Equation~\ref{boltzfull} with the appropriate expressions for $\Gamma$
(as given by Equations~\ref{mbexts} and \ref{lbexts}) describes the
steady state of self-gravitating collisionless systems. The
relaxation term $\Gamma$ describes the continual evolution of the
coarse-grained distribution function, which is due to the combimation
of the dynamical evolution of the fine-grained DF on microscopic
scales, and coarse-graining.

\acknowledgments The authors gratefully acknowledge support from NASA
Astrophysics Theory Program grant NNX07AG86G.  We would also like to
thank the anonymous referee for numerous constructive comments.

\begin{thebibliography}{99}

\bibitem[Binney \& Tremaine(1987)]{bt87}
Binney, J., Tremaine, S. 1987, Galactic Dynamics, (Princeton,
NJ:Princeton)

\bibitem[Bordel(2010)]{b10}
Bordel, S. 2010, Physica A, 389, 4564

\bibitem[de Groot \& Mazur(1984)]{dgm84}
de Groot, S. R., Mazur, P. 1984, Non-Equilibrium Thermodynamics,
(Mineola, NY:Dover)

\bibitem[Grandy(2008)]{g08}
Grandy, W. T. 2008, Entropy and the Time Evolution of Macroscopic
Systems, (New York, NY:Oxford)

\bibitem[Hjorth \& Williams(2010)]{hw10}
Hjorth, J., Williams, L. L. R. 2010, \apj, 722, 851

\bibitem[Kandrup(1998)]{k98}
Kandrup, H. E. 1998, \apj, 500, 120

\bibitem[Lynden-Bell(1967)]{lb67}
Lynden-Bell, D. 1967, \mnras, 136, 101

\bibitem[Madsen(1996)]{m96}
Madsen, J. 1996, \mnras, 280, 1089

\bibitem[Martyushev \& Seleznev(2006)]{ms06}
Martyushev, L. M., Seleznev, V. D. 2006, Phys. Rep. 426, 1

\bibitem[Merritt \& Aguilar(1985)]{ma85}
Merritt, D., Aguilar, L. 1985, \mnras, 217, 787

\bibitem[Moore \etal(1998)]{m98}
Moore, B., Governato, F., Quinn, T., Stadel, J., Lake, G. 1998,
\apj, 499, L5

\bibitem[Navarro, Frenk, \& White(1996)]{nfw96}
Navarro, J. F., Frenk, C. S., White, S. D. M. 1996, \apj, 462, 563

\bibitem[Plastino \& Plastino(1993)]{pp93}
Plastino, A. R., Plastino, A. 1993, Phys. Lett. A, 174, 384

\bibitem[Prigogine(1978)]{p78}
Prigogine, I. 1978, Science 201, 777

\bibitem[Prigogine \& Geheniau(1986)]{pg86}
Prigogine. I., Geheniau, J. 1986, Proc. Natl. Acad. Sci. USA, 83, 6245

\bibitem[Taylor \& Navarro(2001)]{tn01}
Taylor, J. E., Navarro, J. F. 2001, \apj, 563, 483

\bibitem[Tremaine, H\'{e}non, \& Lynden-Bell(1986)]{t86}
Tremaine, S., H\'{e}non, M., Lynden-Bell, D. 1986, \mnras, 219, 285

\bibitem[Tsallis(1988)]{ts88}
Tsallis, C. 1988, J. Stat. Phys., 52, 479

\bibitem[van Albada(1982)]{va82}
van Albada, T. S. 1982, \mnras, 201, 939

\bibitem[Ziegler(1961)]{z61}
Ziegler, H. 1961, Ing. Arch. 30, 410

\end{thebibliography}
\end{document}